\documentclass[conference]{IEEEtran}
\usepackage{amsmath,amssymb,amsthm}
\usepackage{graphicx}
\usepackage{cite}
\usepackage{xcolor}
\usepackage{hyperref}

\usepackage{mathtools}
\mathtoolsset{showonlyrefs}

\newcommand{\Asf}{\mathsf{A}}
\newcommand{\Ksf}{\mathsf{K}}
\newcommand{\Nsf}{\mathsf{N}}
\newcommand{\Msf}{\mathsf{M}}
\newcommand{\Rsf}{\mathsf{R}}
\newcommand{\Bsf}{\mathsf{B}}
\newcommand{\qsf}{\mathsf{q}}
\newcommand{\rsf}{\mathsf{r}}
\newcommand{\asf}{\mathsf{a}}
\newcommand{\FF}{\mathbb{F}}
\newcommand{\Ic}{\mathcal{I}}
\newcommand{\Wc}{\mathcal{W}}
\newcommand{\Sc}{\mathcal{S}}

\newcommand{\Qc}{\mathcal{Q}}
\newcommand{\Ec}{\mathcal{E}}
\newcommand{\Gc}{\mathcal{G}}
\newcommand{\Tc}{\mathcal{T}}
\newcommand{\Bc}{\mathcal{B}}
\newcommand{\Hc}{\mathcal{H}}
\newcommand{\dv}{\mathbf{d}}
\newcommand{\ev}{\mathbf{e}}
\newcommand{\uv}{\mathbf{u}}
\newcommand{\gv}{\mathbf{g}}
\newcommand{\Gm}{\mathbf{G}}
\newcommand{\row}{{\rm row}}

\newtheorem{thm}{Theorem}
\newtheorem{cor}{Corollary}
\newtheorem{lem}{Lemma}
\theoremstyle{definition}

\theoremstyle{remark}
\newtheorem{rem}{Remark}

\allowdisplaybreaks

\begin{document}

\title{Hotplug Coded Caching: Uncoded versus Row-Wise Coded Placements}

\author{\IEEEauthorblockN{Daniela Tuninetti}
\IEEEauthorblockA{University of Illinois Chicago, Chicago, IL 60607, USA;
Email: danielat@uic.edu}}

\maketitle

\begin{abstract}
In the hotplug coded caching model, $K$ users fill their caches of size $M$ as a function of $N$ files, but only an unknown subset of $A \leq K$ users is active in the delivery phase. In all regimes where the optimal hotplug tradeoff is known, it depends only on $A$, not on $K$; whether this holds in general is open. We show that:
(i) Under uncoded placement, YMA delivery with MAN placement, where offline users are assigned the demands of active users, is optimal. Hence, offline users cause a strict load penalty at intermediate memory sizes, while the MAN placement is optimal for every number of active users.
(ii) Under row-wise placement (RWP), where each cache is a function of the user’s own row of each file, every classical scheme can be embedded in a Reed-Solomon code. The resulting hotplug scheme has the tradeoff of the classical model with $A$ users, for every  $A \leq K$: there is no load penalty, but the placement depends on $K$.
We then derive a converse for RWP, which shows that: 
(a) for $2 \leq N \leq A$, the scheme of Gómez-Vilardebó (GV) is RWP optimal for $AM \leq N/(N-1)$, which settles the RWP tradeoff for $N = 2$; 
(b) YMA is RWP optimal at $AM = N$; and 
(c) for $N = A = 3$, GV is RWP optimal for $M \leq 1$. 
Hence, the Reed-Solomon embeddings of these schemes are optimal among embedded RWP schemes.
\end{abstract}


\section{Introduction}
Classical coded caching trades off network load for storage at the end users. Maddah-Ali and Niesen~\cite{MAN} introduced a shared-link model with cache-aided users, and proposed a scheme with an uncoded combinatorial placement and a coded delivery (referred to as MAN), which is to within a constant factor of a cut-set bound.
Let $\Ksf$ be the number of users, $\Nsf$ the number of files, and $\Msf\in[0,\Nsf]$ the cache size.
Under the constraint of uncoded placement, the exact worst-case memory-load tradeoff was characterized by Wan \emph{et al.}~\cite{WTP} for $\Nsf\geq\Ksf$, through an acyclic index coding converse, and by Yu \emph{et al.}~\cite{YMA} for all $\Nsf$ and $\Ksf$; the YMA scheme of~\cite{YMA}  is also optimal to within a factor of two without restrictions on the placement~\cite{YMAfactor2}. Coded placement strictly improves on uncoded placement at small memory sizes: the scheme of Chen, Fan and Letaief (CFL)~\cite{chen2016fundamental} is optimal for $\Ksf\Msf\leq1$ when $\Nsf\leq\Ksf$, and the scheme of Gómez-Vilardebó (GV)~\cite{gomez2018fundamental} is optimal for $\Ksf\Msf\leq \Nsf/(\Nsf-1)$ when $\lceil(\Ksf+1)/2\rceil\leq\Nsf\leq\Ksf$~\cite{kumar2021exact}. Linear coded placement schemes based on rank-metric codes were developed in~\cite{tian2018caching,zhang2020thesis}, which outperform YMA and GV in some regimes;~\cite{tian2018symmetry,caoxu2020} derived computer-aided converse bounds.

The classical model requires every user that took part in the placement phase to also take part in the delivery phase. In~\cite{MT2022offline} we introduced the \emph{hotplug} model, in which only $\Asf$ of the $\Ksf$ users are active in the delivery phase, and the identity of the active users is unknown at the time of placement. In~\cite{MT2022offline,MT2024arxiv} we proposed schemes based on Maximum Distance Separable (MDS) coded placement that are exactly optimal in the small and large memory regimes, and optimal to within a factor of two otherwise. In all cases where we could establish exact optimality, the optimal hotplug tradeoff depends only on the number of active users $\Asf$, and not on the total number of users $\Ksf$; whether this holds in general was left open.
Hotplug placement delivery arrays were proposed in~\cite{rajput2024improved,chinnapadamala2024new}; their algorithms provide examples of tradeoffs that outperform the baseline scheme in~\cite{MT2022offline}, but no closed-forms are available.

Models of random users' inactivity have also been proposed~\cite{liao2022inactivity,YaparWanSchaeferCaireTCOM2019}, which cannot be compared to hotplug. Decentralized placement~\cite{MANdecentralized} is compatible with hotplug, since the placement does not depend on which users, or how many, will be active; the decentralized load is the limit of the centralized load when the number of users tends to infinity.

\subsection{Main Contributions}
\emph{(1) Uncoded placement.} We show that the extension of YMA in~\cite{MT2022offline} (referred to as YMA+), in which the offline users are assigned the demands of the active users, is optimal under uncoded placement (Theorem~\ref{thm:uncoded}). Hence, under uncoded placement and when $\min(\Asf,\Nsf)\geq2$, a system with $\Asf<\Ksf$ active users pays a strict load penalty at intermediate memory sizes with respect to a classical system with $\Asf$ users (Corollary~\ref{cor:penalty}). Moreover, the MAN placement is universal: it is optimal for every number of active users (Corollary~\ref{cor:unknownA}). The penalty is thus in the load only: the active users do not see the symmetric MAN placement structure of $\Asf$ users, but the restriction of a $\Ksf$-user structure to an unknown subset. Hence, matching the classical load with $\Asf$ users requires coded placement.

\emph{(2) Reed--Solomon embedding of RWP placement (RWP).} In an RWP for a classical system (i.e., $\Ksf=\Asf$), each file is a matrix with $\Asf$ rows, and the cache of a user is a function of its row of every file; the caches are thus independent and $\Msf\leq\Nsf/\Asf$. We show that every classical RWP scheme can be embedded in a Reed--Solomon code, which yields a hotplug scheme for every $\Ksf\geq\Asf$ whose memory-load tradeoff is the same as that of the classical model (Theorem~\ref{thm:RS}), i.e., there is no load penalty for offline users. This greatly generalizes some schemes of~\cite{MT2022offline} and gives a tradeoff complementary to that of uncoded placement: the placement  depends on both $\Asf$ and $\Ksf$, but the load depends only on $\Asf$.

\emph{(3) Converse for RWP placement.} We then ask how good RWP is. We derive a converse bound for RWP in the classical model (Theorem~\ref{thm:chain}), in the spirit of the acyclic index coding bound for uncoded placement~\cite{WTP}. It shows that, for every $2\leq\Nsf\leq\Asf$, the GV scheme is optimal among RWP schemes for $\Asf\Msf\leq1+1/(\Nsf-1)$; thus the RWP tradeoff is fully characterized for $\Nsf=2$. It also shows that the YMA point with one cached subfile per file, at $\Asf\Msf=\Nsf$, is RWP optimal for every $\Nsf$ and $\Asf$ (Corollary~\ref{cor:tight}).
For $\Nsf=\Asf=3$, a separate converse shows that the GV scheme is optimal among RWP schemes for $\Asf\Msf\leq\Nsf$ (Theorem~\ref{thm:N=A=3}).
Consequently, the Reed--Solomon embeddings of these schemes are optimal among embedded RWP schemes for every $\Ksf\geq\Asf$ (Corollary~\ref{cor:hotplugrow}).

\subsection{Notation}
Calligraphic symbols denote sets, bold symbols vectors and matrices, and sans-serif symbols system parameters.
$\binom{a}{b}=0$ unless $a\geq b\geq0$.
For integers $a\leq b$, $[a:b]:=\{a,\ldots,b\}$ and $[b]:=[1:b]$.
For a set $\Gc$, $\Omega_\Gc^t:=\{\Tc\subseteq\Gc: |\Tc|=t\}$.
For a set $\Ic\subseteq\mathbb{N}$, $X_\Ic:=(X_n : n\in \Ic)$,
and $\dv[\Ic]$ is the sub-vector of $\dv$ restricted by $\Ic$.

\section{Problem Setting}
\label{sec:model}

An $(\Asf,\Ksf,\Nsf)$ hotplug model consists of a server with $\Nsf$ files and $\Ksf$ cache-aided users connected to the server via a shared link. File $F_n$, $n\in[\Nsf]$, contains $\Bsf$ i.i.d. uniformly distributed symbols over $\FF_\qsf$, for a prime power $\qsf$.
In the placement phase, the server fills $Z_k$, the cache of user $k\in[\Ksf]$, as
\begin{align}
H(Z_k |  F_{[\Nsf]})=0,
\quad
H(Z_k)\leq\Msf.
\label{eq:placement}
\end{align}
In the delivery phase, a set $\Ic\in\Omega^{\Asf}_{[\Ksf]}$ of users is active, unknown at the time of placement; active user $k\in\Ic$ requests file $d_k\in[\Nsf]$. The server sends $X$ as
\begin{align}
H(X |  F_{[\Nsf]},\Ic,\dv[\Ic])=0,
\quad
H(X)\leq\Rsf.
\label{eq:delivery}
\end{align}
Every active user decodes its demanded file as
\begin{align}
H(F_{d_k} |  X,Z_k,\Ic,\dv[\Ic])=0,\quad\forall k\in\Ic.
\label{eq:decoding}
\end{align}
The optimal worst-case load $\Rsf^\star_{(\Asf,\Ksf,\Nsf)}(\Msf)$ is the infimum of the achievable $\Rsf$ over all $\qsf$, $\Bsf$ and schemes that satisfy~\eqref{eq:placement},~\eqref{eq:delivery} and~\eqref{eq:decoding} for every $\Ic$ and $\dv[\Ic]$. The case $\Asf=\Ksf$ is the classical MAN model with $\Asf$ users. We let $\rsf:=\min(\Asf,\Nsf)$.

\paragraph*{Uncoded placement} If every cache is a subset of the symbols of the files, the placement is said to be uncoded, and the optimal worst-case load is $\Rsf^\star_{{\rm u}(\Asf,\Ksf,\Nsf)}(\Msf)\geq\Rsf^\star_{(\Asf,\Ksf,\Nsf)}(\Msf)$.

\paragraph*{RWP placement} For a classical system with $\Asf$ users, a placement is \emph{RWP} if each file is split into $\Asf$ rows, $F_n=(F_{n,1},\ldots,F_{n,\Asf})$, $n\in[\Nsf]$, and the cache of user $k\in[\Asf]$ is a function of the $k$-th row of every file, i.e.,
$
Z_k=\phi(F_{1,k},\ldots,F_{\Nsf,k}),
%\label{eq:rowwise}
$
where using the same function $\phi$ at every user is without loss of generality~\cite{tian2018symmetry}. 
Since a row has entropy $\Nsf/\Asf$, RWP is restrited to $\Asf\Msf\leq\Nsf$.
The optimal RWP load is denoted by $\Rsf^\star_{\row(\Asf,\Nsf)}(\Msf)\geq\Rsf^\star_{(\Asf,\Asf,\Nsf)}(\Msf)$.
RWP may be uncoded, and uncoded placement can be RWP, thus, no general ordering between $\Rsf^\star_{\row(\Asf,\Nsf)}$ and $\Rsf^\star_{{\rm u}(\Asf,\Asf,\Nsf)}$ exists.

\section{Known Results}
\label{sec:prior}

\subsection{YMA+ (uncoded) scheme~\cite{MT2022offline} for hotplug}
For $\Ksf$ users and $t\in[0:\Ksf]$, the MAN placement splits $F_n=(F_{n,\Wc}:\Wc\in\Omega^t_{[\Ksf]})$, $n\in[\Nsf]$, and user $k\in[\Ksf]$ stores $F_{n,\Wc}$ for all $n$ and $\Wc\ni k$; hence, $\Msf=\Nsf t/\Ksf$. For $\Asf<\Ksf$, YMA+~\cite{MT2022offline} assigns to offline users the demand of an active user and runs the YMA delivery~\cite{YMA} for all $\Ksf$ users, with at most $\rsf$ distinct demands. YMA+ thus achieves the lower convex envelope of
\begin{align}
(\Msf_t,\Rsf_t) =\Big(\frac{\Nsf t}{\Ksf}, c_t\Big),
\ c_t:=\frac{\binom{\Ksf}{t+1}-\binom{\Ksf-\rsf}{t+1}}{\binom{\Ksf}{t}},
\ t\in[0:\Ksf],
\label{eq:YMAplus}
\end{align}
denoted by $\Rsf^{(\rm YMA+)}_{(\Asf,\Ksf,\Nsf)}(\Msf)$. For $\Asf=\Ksf$ this is YMA, which is optimal under uncoded placement~\cite{YMA,WTP}. For every set of active users, the caches of the active users and the delivery form a classical scheme for $\Asf$ users, hence
\begin{align}
\frac{\Rsf^{(\rm dec)}_{(\Asf,\Nsf)}}{2.00884}
\stackrel{\rm(i)}{\leq}
    \Rsf^\star_{(\Asf,\Asf,\Nsf)}
\leq\Rsf^\star_{(\Asf,\Ksf,\Nsf)}
\leq\Rsf^{(\rm YMA+)}_{(\Asf,\Ksf,\Nsf)}
\stackrel{\rm(ii)}{\leq}
     \Rsf^{(\rm dec)}_{(\Asf,\Nsf)},
\label{eq:bounds}
\end{align}
where the lower bounds depend only on $\Asf$, and the inequalities marked as (i) and (ii) are from~\cite{YMAfactor2}, with $\mu := \Msf/\Nsf$
\begin{align}
\Rsf^{(\rm dec)}_{(\Asf,\Nsf)} = \lim_{\Ksf\to\infty}\Rsf^{(\rm YMA+)}_{(\Asf,\Ksf,\Nsf)}
= \frac{1-\mu}{\mu}\big(1-(1-\mu)^\rsf\big), 
\end{align}
being the load with decentralized placement~\cite{MANdecentralized,YMA}.
The chain of inequalities in~\eqref{eq:bounds} shows that YMA+ is optimal to within a factor of two~\cite{MT2024arxiv}.

\subsection{GV (row-wise) scheme~\cite{gomez2018fundamental} for classical coded caching}
The GV scheme~\cite{gomez2018fundamental} for $\Asf$ users is an RWP scheme that further splits each row into pieces indexed by the $g$-subsets of $[\Nsf]$ and caches sums of pieces of $g$ distinct files of the same row; it achieves the lower convex envelope of $(0,\rsf)$ and the points
\begin{align}
(\Msf_g,\Rsf_g) =\Big(\frac{\Nsf}{\Asf g},\ \rsf-\frac{\Nsf(\rsf+1)}{\Asf(g+1)}\frac{\binom{\rsf}{g}}{\binom{\Nsf}{g}}\Big),\quad g\in[\Nsf].
\label{eq:GV}
\end{align}
For $g=\Nsf\leq\Asf$, this is the CFL point $(\frac1\Asf,\Nsf(1-\frac1\Asf))$~\cite{chen2016fundamental}, which caches $\sum_nF_{n,k}$.
For $g=1$, this is the YMA point with $t=1$, $(\frac\Nsf\Asf,\rsf-\frac{\rsf(\rsf+1)}{2\Asf})$, which caches the row itself.
The line through $(0,\Nsf)$ and the CFL point is optimal for $\Nsf\leq\Asf$~\cite{chen2016fundamental}.
The line through the CFL point and the GV point with $g=\Nsf-1$ is optimal for $\Nsf=\Asf$~\cite{gomez2018fundamental} and for $\lceil(\Asf+1)/2\rceil\leq\Nsf\leq\Asf$~\cite{kumar2021exact}.

\subsection{MDS-coded placement~\cite{MT2022offline,MT2024arxiv} for hotplug}
Our past work showed that the following points of a classical system with $\Asf$ users are achievable in the hotplug model for every $\Ksf\geq\Asf$: the CFL point for $\Asf\geq\Nsf$~\cite[Scheme~new2]{MT2022offline} (HT2 in~\cite{MT2024arxiv}), the YMA point with $t=1$~\cite[Scheme~new1]{MT2022offline} (HT1 with $t=1$ in~\cite{MT2024arxiv}), and the YMA point with $t=\Asf-1$ for $\Asf\geq3$~\cite[Scheme~HT3]{MT2024arxiv}. Together with~\eqref{eq:bounds}, these results give $\Rsf^\star_{(\Asf,\Ksf,\Nsf)}=\Rsf^\star_{(\Asf,\Asf,\Nsf)}$ wherever the latter is attained by one of these points. In~\cite[Scheme~new1]{MT2022offline} we also added an MDS layer to the MAN placement with $t\geq2$, which attains the YMA load of a classical system with $\Asf$ users, but with a memory size that depends on both $\Asf$ and $\Ksf$.

\section{Main Results}
\label{sec:main}
We state next our main results, with proof sketches; the full proofs can be found in Appendix. 
%~\cite{thislonger}.

\subsection{Uncoded placement for hotplug}
\begin{thm}[Optimality of YMA+]
\label{thm:uncoded}
For an $(\Asf,\Ksf,\Nsf)$ hotplug model, $\Rsf^\star_{{\rm u}(\Asf,\Ksf,\Nsf)}(\Msf)=\Rsf^{(\rm YMA+)}_{(\Asf,\Ksf,\Nsf)}(\Msf)$ for all $\Msf\in[0,\Nsf]$.
\end{thm}
{\it Proof Sketch:}
For the converse, we apply the acyclic index coding bound of~\cite{WTP} to the active users only and average it over the set of active users, their ordering and their distinct demands, which gives $\Rsf\geq\sum_s x_s c_s$, where $x_s$ is the fraction of the library cached by exactly $s$ of the $\Ksf$ users; minimizing over the ditribution $\{x_s\}$ subject to $\sum_s s\,x_s\leq\Ksf\Msf/\Nsf$, as in~\cite{YMA,WTP}, gives the YMA+ load.
\hfill$\square$

\begin{cor}[Penalty of offline users]
\label{cor:penalty}
For an $(\Asf,\Ksf,\Nsf)$ hotplug model under uncoded placement, we have:
\\
(C\ref{cor:penalty}.I) if $\rsf\geq2$ and $\Ksf>\Asf$, then for every $t\in[\Asf-1]$
\begin{align}
\Rsf^\star_{{\rm u}(\Asf,\Ksf,\Nsf)}\Big(\frac{\Nsf t}{\Asf}\Big)>\Rsf^\star_{{\rm u}(\Asf,\Asf,\Nsf)}\Big(\frac{\Nsf t}{\Asf}\Big);
\label{eq:strictgap}
\end{align}
(C\ref{cor:penalty}.II) $\Rsf^\star_{{\rm u}(\Asf,\Ksf,\Nsf)}(\Msf)=\Rsf^\star_{{\rm u}(\Asf,\Asf,\Nsf)}(\Msf)$ for $\Msf\in[0,\frac\Nsf\Ksf]\cup[\frac{\Nsf(\Ksf-1)}{\Ksf},\Nsf]$, i.e., the dependance on $\Ksf$ is in the mid-memory regime.
\end{cor}

\begin{cor}[Universality of the MAN placement]
\label{cor:unknownA}
For an $(\Asf,\Ksf,\Nsf)$ hotplug model, 
%let $\Msf=\Nsf t/\Ksf$, $t\in[0:\Ksf]$. T
the MAN placement 
%with parameter $t$, 
followed by YMA+ for the $\asf$ users that turn out to be active, attains $\Rsf^\star_{{\rm u}(\asf,\Ksf,\Nsf)}(\Msf)$ simultaneously for every $\asf\in[\Ksf]$. 
%For other values of $\Msf$, memory sharing between the MAN placements with parameters $\lfloor\Ksf\Msf/\Nsf\rfloor$ and $\lceil\Ksf\Msf/\Nsf\rceil$ has the same property.
\end{cor}

In words, for hotplug under uncoded placement, not knowing \emph{which} users will be active has a load cost (Corollary~\ref{cor:penalty}), whereas not knowing \emph{how many} has none (Corollary~\ref{cor:unknownA}). Since the inequality in~\eqref{eq:strictgap} is strict, an optimality result for the hotplug model that matches the classical lower bound in~\eqref{eq:bounds} necessarily requires coded placement.

\subsection{Reed--Solomon embedding of RWP  for hotplug}
\begin{thm}[Reed--Solomon embedding]
\label{thm:RS}
Let a classical scheme for $\Asf$ users with RWP, over a field $\FF_{\qsf_0}$, achieve $(\Msf,\Rsf)$, with possibly nonlinear cache placement and delivery functions. Then $(\Msf,\Rsf)$ is also achievable in the $(\Asf,\Ksf,\Nsf)$ hotplug model for every $\Ksf\geq\Asf$, over any extension field $\FF_\qsf$ of $\FF_{\qsf_0}$ with $\qsf\geq\Ksf-1$.
\end{thm}
{\it Proof Sketch:}
The main idea is to view each file as a matrix $F_n\in\FF_\qsf^{\Asf\times\Bsf/\Asf}$, $n\in[\Nsf]$. Let $\ev_a\in\FF_\qsf^{1\times\Asf}$ be the standard basis vector with a one in position $a\in[\Asf]$. In the classical model with RWP, the cache contents are $Z_a=\phi(\ev_aF_1,\ldots,\ev_aF_\Nsf)$, $a\in[\Asf]$. In hotplug, users are assigned the ``MDS vectors''
\begin{align}
\gv_k:=[1,\alpha_k,\ldots,\alpha_k^{\Asf-1}],\ k\in[\Ksf-1],\ \gv_\Ksf:=[0,\ldots,0,1],
\label{eq:ex g}
\end{align}
where $\alpha_1,\ldots,\alpha_{\Ksf-1}\in\FF_\qsf$ are distinct, which requires $\qsf\geq\Ksf-1$; users cache $Z_k=\phi(\gv_kF_1,\ldots,\gv_kF_\Nsf)$, $k\in[\Ksf]$, thus same cache size as in the classical model.
Any $\Asf$ of the vectors in~\eqref{eq:ex g} are linearly independent (Vandermonde). Once the set of active users $\Ic=\{i_1,\ldots,i_\Asf\}$ is known, let $\Gm_\Ic$ be the invertible matrix with rows $\gv_{i_1},\ldots,\gv_{i_\Asf}$, and define the library $F'_n:=\Gm_\Ic F_n$, $n\in[\Nsf]$. Since $\ev_jF'_n=\gv_{i_j}F_n$, the cache $Z_{i_j}$ is the cache of classical user $j$ for the library $(F'_n)_n$. The server then uses the classical transmission for the library $(F'_n)_n$ and demands $(d_{i_j})_{j\in[\Asf]}$; user $i_j$ decodes $F'_{d_{i_j}}$ and recovers $F_{d_{i_j}}=\Gm_\Ic^{-1}F'_{d_{i_j}}$.
\hfill$\square$

\begin{rem}
\label{rem:complementary}
Under uncoded placement, the optimal placement does not depend on the number of active users (Corollary~\ref{cor:unknownA}), but the load depends on $\Ksf$ (Corollary~\ref{cor:penalty}). The Reed--Solomon (RS) embedding has the opposite behavior: the load depends only on $\Asf$, but the placement function depends on $\Asf$ (the number of rows, i.e., the dimension of the RS code) and on $\Ksf$ (the length of the RS code). If only $\asf<\Asf$ users turn out to be active, the server can still deliver to any $\Asf$-subset of users that contains them, with load $\Rsf$, but this is in general not optimal.
\end{rem}

\begin{rem}
Theorem~\ref{thm:RS} applied to the GV scheme in~\eqref{eq:GV} gives the achievability of the GV points of a classical system with $\Asf$ users, for every $\Ksf\geq\Asf$.
It recovers HT2 and HT1 with $t=1$ of~\cite{MT2024arxiv};
%(GV with $g=\Nsf$ and $g=1$, respectively); 
HT1 with $t\geq2$ and HT3 are not RWP. %RWP, since their memory size exceeds $\Nsf/\Asf$.
Theorem~\ref{thm:RS} is more general than~\cite{MT2024arxiv}: (a) it applies to every RWP scheme, with a linear or nonlinear placement and delivery; and (b) any RWP scheme that improves on GV in the open range of Corollary~\ref{cor:tight} immediately yields an improved hotplug scheme in the same regime.
For example, in the scheme of Amiri and Gündüz~\cite{amiri2016fundamental}, user $k$ caches $F_{n,k}+F_{n+1,k}$, $n\in[\Nsf-1]$, at $\Msf=(\Nsf-1)/\Asf$; this is an RWP, hence it yields a hotplug scheme with the same load for every $\Ksf\geq\Asf$. In this case the embedding gives no new point, since the scheme of~\cite{amiri2016fundamental} is strictly dominated by GV for $\Nsf\geq3$, and it coincides with CFL for $\Nsf=2$.
\end{rem}

\subsection{Converse for RWP in the classical model}
Motivated by the ability to embed any classical RWP scheme in the hotplug model for any $\Ksf \geq \Asf$, we next study the fundamental limits of RWP for classical coded caching with $\Asf$ users.
For such a classical system, let
$h_s:=H(Z_k |  F_\Sc)$ for $|\Sc|=s\in[0:\Nsf]$; by the symmetry argument in~\cite{tian2018symmetry}, without loss of generality $H(Z_k |  F_\Sc)$ does not depend on the user $k\in[\Asf]$ or on the subset $\Sc\in\Omega_{[\Nsf]}^s$. For every placement, $h_s$ is non-increasing in $s$ (conditioning reduces entropy) and concave in $s$ (submodularity of entropy), and
\begin{align}
0=h_\Nsf\leq h_{\Nsf-1}\leq\cdots\leq h_1\leq h_0=\Msf,
\label{eq:lem2}
\end{align}
where, without loss of generality, $h_0=\Msf$: otherwise the unused memory can be filled with uncoded symbols of the user's row, which preserves the RWP structure and does not increase the load.

\begin{thm}[RWP converse]
\label{thm:chain}
For an $(\Asf,\Nsf)$ classical coded caching model with $\Msf\leq\Nsf/\Asf$,
\begin{align}
\Rsf^\star_{\row(\Asf,\Nsf)}(\Msf)\geq\min_{h\in\Hc(\Msf)}\ \max_{L\in[\rsf]}\Big[L(1+h_L)-\sum_{i=0}^{L-1}h_i\Big],
\label{eq:chain}
\end{align}
where $\Hc(\Msf)$ is the set of concave $(h_s)$ that satisfy~\eqref{eq:lem2} and
\begin{align}
\max\Big(0, h_0-\frac{s}{\Asf}\Big)\leq h_s\leq\min\Big(h_0,\frac{\Nsf-s}{\Asf}\Big).
\label{eq:lem3}
\end{align}
\end{thm}
{\it Proof Sketch:}
Fix $L\in[\rsf]$, distinct users $u_1,\ldots,u_L$ requesting distinct files $d_1,\ldots,d_L$;
let $F_{<i}:=(F_{d_\ell} : \ell<i)$ and $Z_{<i}:=(Z_{u_\ell}: \ell<i)$.
Then, for every $L$
\begin{align}
\Rsf
&\geq H(X)
\geq I(X;F_{<L+1},Z_{<L+1})
\\&= H(F_{<L+1},Z_{<L+1})-H(F_{<L+1},Z_{<L+1} |  X)
\\&= L+Lh_L-H(F_{<L+1},Z_{<L+1} |  X)
\\&\geq L+Lh_L-\sum_{i}H(Z_{u_i} | F_{<i})
= L+Lh_L-\sum_{i}h_{i-1}.
\end{align}
The bound in~\eqref{eq:lem3} follows from the RWP structure.
\hfill$\square$

Theorem~\ref{thm:chain} plays the role of the acyclic index coding bound for uncoded placement in~\cite{WTP}: the conditional cache entropies $h_s$ replace the amounts of data cached by a given number of users, and the minimization over $\Hc(\Msf)$ replaces the optimization over the placement distribution.
For $\Nsf\leq\Asf$, the optimization in~\eqref{eq:chain} has a closed form.

\begin{cor}[Theorem~\ref{thm:chain} for $\Nsf\leq\Asf$]
\label{cor:closed}
For an $(\Asf,\Nsf)$ classical coded caching model with $\Msf\leq\Nsf/\Asf\leq1$,
\begin{align}
\Rsf^\star_{\row(\Asf,\Nsf)}(\Msf)\geq\Nsf-\frac{1}{\Asf}\sum_{j\in[\Nsf]}\min(\Asf\Msf,\,j),
\label{eq:closed}
\end{align}
i.e, piecewise linear among $\big(\frac{q}{\Asf},\ \Nsf-\frac{q(2\Nsf-q+1)}{2\Asf}\big)$, $q\in[0:\Nsf]$.
\end{cor}

\begin{rem}
For $\Nsf>\Asf$, \eqref{eq:chain} is a linear program in $(h_1,\ldots,h_{\Nsf-1})$ that does not appear to have a simple closed-form solution, but it is easy to solve numerically.
\end{rem}

\begin{cor}[RWP optimality]
\label{cor:tight}
For an $(\Asf,\Nsf)$ classical model we have:
(C\ref{cor:tight}.I) If $2\leq\Nsf\leq\Asf$,
\begin{align}
\Rsf^\star_{\row(\Asf,\Nsf)}(\Msf)=
\begin{cases}
\Nsf(1-\Msf), & 0\leq\Msf\leq\frac1\Asf,\\
\Nsf-\frac{1}{\Asf}-(\Nsf-1)\Msf, & \frac1\Asf\leq\Msf\leq\frac{\Nsf}{\Asf(\Nsf-1)},
\end{cases}
\end{align}
i.e., the GV scheme is RWP optimal on the segment between the points with $g=\Nsf$ and $g=\Nsf-1$, for every $\Nsf\leq\Asf$.
\\
(C\ref{cor:tight}.II) For $\Nsf=2\leq\Asf$ and $\Msf\in[0,\frac2\Asf]$,
\begin{align}
\Rsf^\star_{\row(\Asf,2)}(\Msf)=\max\{2-2\Msf,\ 2-\tfrac1\Asf-\Msf\}.
\end{align}
(C\ref{cor:tight}.III) For $\Asf\Msf\leq\Nsf$, and every $\Nsf$ and $\Asf$,
\begin{align}
\Rsf^\star_{\row(\Asf,\Nsf)}(\Msf)\geq\rsf-\frac{\rsf(\rsf+1)}{2\Asf},
\label{eq:flat}
\end{align}
with equality at $\Asf\Msf=\Nsf$, attained by YMA with $t=1$.
\end{cor}

Corollary~\ref{cor:tight} has the following consequences:
\begin{itemize}
\item (C\ref{cor:tight}.I) extends the optimality of the GV point with $g=\Nsf-1$, known among all placements for $\lceil(\Asf+1)/2\rceil\leq\Nsf\leq\Asf$~\cite{kumar2021exact}, to every $2\leq\Nsf\leq\Asf$ among RWP schemes. Its optimality without placement restrictions for $\Nsf<\lceil(\Asf+1)/2\rceil$ remains open.

\item For $\Nsf=2$, the tradeoff without restrictions on the placement is known for $\Asf\leq3$~\cite{MAN,tian2018symmetry}, while for $\Asf\geq4$ it is open on $(\frac1\Asf,\,2(1-\frac2\Asf))$, except that for $\Asf=4$ it is known on $[\frac23,2]$~\cite{tian2018caching,tian2018symmetry}; in all cases the open range contains $(\frac1\Asf,\frac2\Asf]$.
(C\ref{cor:tight}.II) characterizes the optimal RWP tradeoff; for its optimality without placement restrictions, see Section~\ref{sec:examples}.

\item (C\ref{cor:tight}.III) holds for every $\Nsf$ and $\Asf$, i.e., also for $\Nsf>\Asf$.

\item
The corner points of~\eqref{eq:closed} are at $\Msf=\frac q\Asf$, $q\in[0:\Nsf]$, while the GV points are at $\Msf=\frac{\Nsf}{\Asf g}$, $g\in[\Nsf]$; hence a GV point has the same memory size as a corner point of~\eqref{eq:closed} if and only if $qg=\Nsf$. In this case, a direct computation gives they have teh same load
%\begin{align}
%\Rsf_g-\Big(\Nsf-\frac{q(2\Nsf-q+1)}{2\Asf}\Big)=\frac{q(q-1)(g-1)}{2\Asf(g+1)},
%\end{align}
%which is zero 
if and only if $g=\Nsf$ (the CFL point, $q=1$) or $g=1$ (the YMA point with $t=1$, $q=\Nsf$). For example, for $\Nsf=4$ and $g=q=2$, the GV point lies $\frac{1}{3\Asf}$ above the converse. The GV point with $g=\Nsf-1$ is RWP optimal even though, for $\Nsf\geq3$, its memory size is not a corner point of~\eqref{eq:closed}: it lies on the segment of~\eqref{eq:closed} with $q=1$.

\item For $3\leq\Nsf\leq\Asf$, the right-hand side of~\eqref{eq:closed} lies strictly below the lower convex envelope of the GV points for $\frac{\Nsf}{\Nsf-1}<\Asf\Msf<\Nsf$, where the RWP optimal load remains open in general; the case $\Nsf=\Asf=3$ is settled by Theorem~\ref{thm:N=A=3} next.
\end{itemize}

\begin{thm}[RWP optimality for $\Nsf=\Asf=3$]
\label{thm:N=A=3}
For $\Nsf=\Asf=3$ and $\Msf\in[0,1]$,
\begin{align}
\Rsf^\star_{\row(3,3)}(\Msf)=\max\Big\{3-3\Msf,\ \tfrac83-2\Msf,\ \tfrac73-\tfrac43\Msf\Big\},
\label{eq:N=A=3}
\end{align}
i.e., the GV scheme is optimal among RWP schemes, linear or not; its corner points are $(0,3)$, $(\frac13,2)$, $(\frac12,\frac53)$ and $(1,1)$. In particular, no RWP scheme attains $(\frac35,\frac32)$.
\end{thm}
{\it Proof Sketch:}
%Achievability is by the GV scheme. 
The first two terms in~\eqref{eq:N=A=3} follow from Corollary~\ref{cor:closed}, which on $[0,1]$ coincides with the outer bound of Tian~\cite{tian2018symmetry} for $(N,K)=(3,3)$, valid for every placement. The third term, $3\Rsf+4\Msf\geq7$, is a separate RWP converse, strictly tighter than Tian's bound on $(\frac12,1)$; it follows by summing Shannon inequalities that involve only $X$, one cache, whole files, and one row of at most two files, together with decoding, the RWP identities, and user--file symmetry.
Since $3\cdot\frac32+4\cdot\frac35=\frac{69}{10}<7$, the linear scheme of Cao and Xu~\cite{caoxu2020} at $(\frac35,\frac32)$, which is optimal among linear schemes, is not RWP, and no RWP scheme attains the same point\footnote{The corner point $(\frac23,\frac43)$ of Tian's bound is not achievable by linear codes~\cite{tian2018symmetry}, nor by any code, since every scheme satisfies $41\Msf+31\Rsf\geq69$~\cite{caoxu2020}. Hence either the Cao--Xu tradeoff is optimal, which would require a tighter outer bound to prove, or it can be improved only by a nonlinear scheme whose placement is not RWP. At $\Msf=\frac23$, for example, the optimal load is only known to lie in $[\frac{125}{93},\frac{17}{12}]$, whereas the RWP optimal load is $\frac{13}{9}$.}.
\hfill$\square$

\subsection{Optimality of RWP in the hotplug model}
\begin{cor}[Optimal Reed--Solomon embedded schemes]
\label{cor:hotplugrow}
Let $\Rsf^\star_{\rm RS(\Asf,\Ksf,\Nsf)}(\Msf)$ be the minimum load over the hotplug schemes obtained from a classical RWP scheme for $\Asf$ users by the embedding of Theorem~\ref{thm:RS}. Then $\Rsf^\star_{\rm RS(\Asf,\Ksf,\Nsf)}(\Msf)=\Rsf^\star_{\row(\Asf,\Nsf)}(\Msf)$ for every $\Ksf\geq\Asf$ and every $\Asf\Msf\leq\Nsf$; in particular, these loads do not depend on $\Ksf$, and they are given by Corollary~\ref{cor:tight} and Theorem~\ref{thm:N=A=3} wherever these characterize $\Rsf^\star_{\row(\Asf,\Nsf)}(\Msf)$.
\end{cor}

\section{Examples}
\label{sec:examples}
We provide two examples to further clarify the relationship between various achievable schemes.

\begin{figure}
\centering
\includegraphics[width=.95\columnwidth]{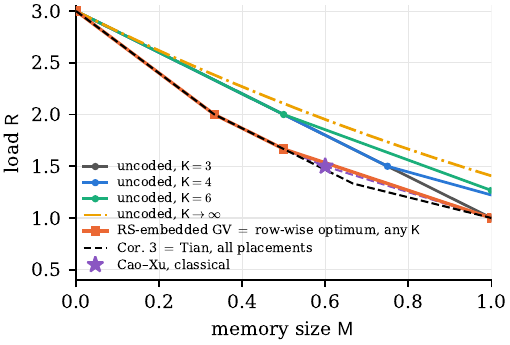}
\caption{Hotplug model with $\Asf=\Nsf=3$: optimal load under uncoded placement for $\Ksf\in\{3,4,6\}$ (Theorem~\ref{thm:uncoded}) and its limit for $\Ksf\to\infty$, i.e., the load of decentralized placement; load of the Reed--Solomon embedding of the GV scheme for $3$ users, which does not depend on $\Ksf$ (Theorem~\ref{thm:RS}) and is RWP optimal (Theorem~\ref{thm:N=A=3}); the point $(\frac35,\frac32)$ of Cao and Xu~\cite{caoxu2020} for the classical system, which is optimal under linear coding, with memory sharing with $(\frac12,\frac53)$ and $(1,1)$; and Corollary~\ref{cor:closed}, which on $[0,1]$ coincides with the converse of~\cite{tian2018symmetry} for every placement}.
\label{fig:example33}
\end{figure}

\paragraph{Case $\Nsf=\Asf=3$}
Fig.~\ref{fig:example33} shows the various tradeoffs in the small memory regime for $\Nsf=\Asf=3$. Under uncoded placement (Theorem~\ref{thm:uncoded}), the optimal load at $\Msf=1$ is $1$ for $\Ksf=3$, $\frac{11}{9}$ for $\Ksf=4$, and $\frac{19}{15}$ for $\Ksf=6$. The Reed--Solomon embedding of the GV scheme for $3$ users (Theorem~\ref{thm:RS}) attains the points $(\frac13,2)$, $(\frac12,\frac53)$ and $(1,1)$ for every $\Ksf\geq3$. By Theorem~\ref{thm:N=A=3}, memory sharing among these points is RWP optimal on $[0,1]$, for every $\Ksf\geq3$. On $[0,1]$, Corollary~\ref{cor:closed} coincides with the converse of Tian~\cite{tian2018symmetry}, which holds for every placement, and it is tight only on $[0,\frac12]$ and at $\Msf=1$.

The optimal tradeoff under {\it linear coding} for the classical system is known~\cite{caoxu2020}: its corner points are at $\Msf\in\{\frac13,\frac12,\frac35,1,2\}$, where $(\frac35,\frac32)$ is attained by the non-RWP scheme of Cao and Xu~\cite{caoxu2020}. With the schemes known to us, the hotplug model matches it for $\Msf\in[0,\frac12]\cup[1,3]$, by Theorem~\ref{thm:RS} and HT3~\cite{MT2024arxiv}.
Whether the optimal load under linear coding of the hotplug model at $\Msf=\frac35$ equals $\frac32$ for $\Ksf\geq4$ is thus open; it hinges on whether this point can be attained in the hotplug model, for example by embedding the scheme of Cao and Xu~\cite{caoxu2020}. By Theorem~\ref{thm:N=A=3}, such an embedding cannot be obtained from a RWP scheme.

\paragraph{Case $\Nsf=2$ and $\Asf\geq 4$}
For the classical model with $\Nsf=2$ files, the optimal tradeoff is known for $\Asf\leq3$~\cite{MAN,tian2018symmetry}, and all its corner points are achievable in the hotplug model for every $\Ksf\geq\Asf$, by Theorem~\ref{thm:RS} and HT3~\cite{MT2024arxiv}.
For $\Asf\geq4$, which is not covered by~\cite{kumar2021exact}, the classical tradeoff is known only for $\Msf\leq\frac1\Asf$~\cite{chen2016fundamental} and $\Msf\geq2(1-\frac2\Asf)$~\cite[Theorem~4.12]{tian2018symmetry}, and it is matched in the hotplug model in both ranges: for $\Msf\leq\frac1\Asf$ by Theorem~\ref{thm:RS}, and for $\Msf\geq2(1-\frac2\Asf)$ by the YMA points of a classical system with $\Asf$ users with $t=\Asf-1$~\cite[Scheme~HT3]{MT2024arxiv} and $t=\Asf-2$\footnote{We can show that, by means of Reed--Muller codes, the YMA scheme can be embedded into the hotplug model for every $t\in[\Asf]$ and every $\Ksf\geq\Asf$.}.

\begin{figure}
\centering
\includegraphics[width=.95\columnwidth]{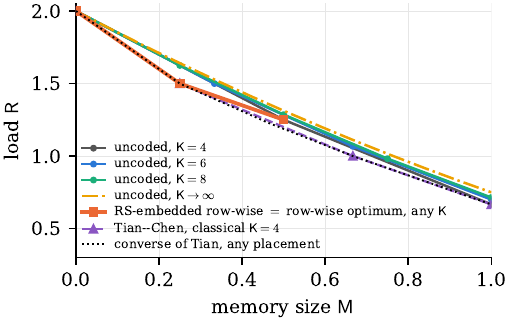}
\caption{Hotplug model with $\Nsf=2$ and $\Asf=4$: optimal load under uncoded placement for $\Ksf\in\{4,6,8\}$ (Theorem~\ref{thm:uncoded}) and its limit for $\Ksf\to\infty$; load of the Reed--Solomon embedding of the RWP schemes for $4$ users, which equals the RWP optimum (Corollary~\ref{cor:tight}) and does not depend on $\Ksf$ (Theorem~\ref{thm:RS}); load of the non-RWP scheme of Tian and Chen~\cite{tian2018caching} for the classical system with $4$ users; and the converse of~\cite{tian2018symmetry} for the classical system, which holds for every placement.}
\label{fig:example24}
\end{figure}

For $\Asf=4$ (see Fig.~\ref{fig:example24}), the scheme of Tian and Chen~\cite{tian2018caching} attains $(\frac23,1)$, which meets the converse of~\cite{tian2018symmetry}; hence the classical tradeoff is known for $\Msf\in[0,\frac14]\cup[\frac23,2]$, whereas, with the schemes known to us, the hotplug model matches it for $\Msf\in[0,\frac14]\cup[1,2]$.
In particular, in the hotplug model at $\Msf=\frac23$ with $\Ksf=3m$, $m\geq2$, YMA+ attains $\Rsf^{(\rm YMA+)}=1+\frac{m-1}{3(3m-1)}$, which increases to $\Rsf^{(\rm dec)}=1+\frac19$ as $m\to\infty$; memory sharing between the embedded YMA points with $t=1$ and $t=2$ attains $\Rsf=1+\frac1{18}$; and memory sharing between the CFL point (Theorem~\ref{thm:RS}) and the embedded YMA point with $t=2$ attains $\Rsf=1+\frac1{27}$, for every $\Ksf\geq4$. Whether $\Rsf^\star_{(4,\Ksf,2)}(\frac23)=1$ for $\Ksf\geq5$ is thus open; it hinges on whether this point can be attained in the hotplug model, for example by embedding the scheme of~\cite{tian2018caching}.

\section{Conclusions}
\label{sec:conclusion}
Under uncoded placement, YMA+ is optimal for the hotplug model: the MAN placement is optimal for every number of active users, and offline users strictly increase the load at intermediate memory sizes. The Reed--Solomon embedding of RWP schemes gives the complementary tradeoff, with a load that depends only on the number of active users. Our converse for RWP establishes the RWP optimality of the GV scheme at its two smallest memory points for every $\Nsf\leq\Asf$, characterizes the RWP tradeoff for $\Nsf=2$ and for $\Nsf=\Asf=3$, and establishes the RWP optimality of the YMA point with $t=1$ for all parameters. Open questions are the RWP tradeoff for $\frac{\Nsf}{\Nsf-1}<\Asf\Msf<\Nsf$ when $\Nsf\geq3$ (except $\Nsf=\Asf=3$), and whether the optimal load of the hotplug model, without any restrictions on the placement, depends only on $\Asf$ and not on $\Ksf$. Our results show that, if it does, the optimal schemes must use coded placement.

\section*{Acknowledgment}
This work was funded in part by NSF under award 2312229.

\section*{Use of AI disclosure}
The Author used Anthropic's Claude (Opus 5.5) during the development and preparation of this manuscript. It assisted in: numerically solving linear programs to search for row-wise schemes and converse bounds, including the extraction of the sparsest Shannon-inequality certificate underlying the proof of Theorem~\ref{thm:N=A=3}; numerically checking the closed-form expressions in the examples; generating the figures; and polishing the text. The Author verified every result, proof, statement and reference in the final manuscript and assumes full responsibility for its contents.

\clearpage
\appendix
\section{Proofs}
\label{sec:proofs}

\subsection{Proof of Theorem~\ref{thm:uncoded}}
Achievability is YMA+ in~\eqref{eq:YMAplus}. By Pascal's rule, $c_s=\sum_{i\in[\rsf]}\binom{\Ksf-s}{i}/\binom{\Ksf}{i}$, and $(c_s)_s$ is non-increasing and convex; hence $\Rsf^{(\rm YMA+)}_{(\Asf,\Ksf,\Nsf)}$ is the piecewise linear interpolation $\hat c(\Ksf\Msf/\Nsf)$ of $(c_s)_s$. For the converse, fix a scheme with uncoded placement and partition each file as $F_n=(F_{n,\Wc}:\Wc\subseteq[\Ksf])$, where $F_{n,\Wc}$ is cached exactly by the users in $\Wc$. Let $x_s:=\frac{1}{\Nsf\Bsf}\sum_{n\in[\Nsf]}\ \sum_{\Wc\in\Omega^s_{[\Ksf]}}|F_{n,\Wc}|$ be the fraction of the library cached by exactly $s$ users; then $\sum_sx_s=1$ and $\sum_ss\,x_s\leq\Ksf\Msf/\Nsf$.

Fix $\Ic$, an ordered list $\uv=(u_1,\ldots,u_\rsf)$ of distinct users in $\Ic$, and demands with distinct $d_{u_1},\ldots,d_{u_\rsf}$. Let $\Qc_i:=(F_{d_{u_i},\Wc}:\Wc\cap\{u_1,\ldots,u_i\}=\emptyset)$, and let $\Ec$ be the rest of the library. Then $Z_{u_i}$ is a function of $(\Ec,\Qc_1,\ldots,\Qc_{i-1})$, and by~\eqref{eq:decoding} and induction on $i$, $H(\Qc_1,\ldots,\Qc_\rsf |  X,\Ec)=0$. Hence
\begin{align}
\Rsf\geq H(\Qc_1,\ldots,\Qc_\rsf)=\sum_{\substack{i\in[\rsf] \\ \Wc:\,\Wc\cap\{u_1,\ldots,u_i\}=\emptyset}}\frac{|F_{d_{u_i},\Wc}|}{\Bsf},
\label{eq:fixed}
\end{align}
the acyclic index coding bound~\cite{WTP} {\it applied to the active users only}. Average~\eqref{eq:fixed} over $\Ic$ uniform in $\Omega^\Asf_{[\Ksf]}$, $\uv$ uniform given $\Ic$, and the distinct demands uniform. Then $\{u_1,\ldots,u_i\}$ is uniform in $\Omega^i_{[\Ksf]}$ and avoids a given $\Wc$ with probability $\binom{\Ksf-|\Wc|}{i}/\binom{\Ksf}{i}$, so that $\Rsf\geq\sum_sx_sc_s\geq\hat c(\sum_ss\,x_s)\geq\hat c(\Ksf\Msf/\Nsf)$ by Jensen's inequality and monotonicity.

\subsection{Proof of Corollary~\ref{cor:penalty}}
(C\ref{cor:penalty}.I): For real $\tau$, let $q^{(\Ksf)}_i(\tau):=\prod_{\ell=0}^{i-1}[\frac{\Ksf-\tau-\ell}{\Ksf-\ell}]^+$; it is convex, non-increasing, and equals $\binom{\Ksf-\tau}{i}/\binom{\Ksf}{i}$ at integers, so $\Rsf^\star_{{\rm u}(\Asf,\Ksf,\Nsf)}(\Msf)\geq\sum_{i\in[\rsf]}q_i^{(\Ksf)}(\Ksf\Msf/\Nsf)$. At $\Msf=\Nsf t/\Asf$, for $\ell\geq1$, $1-\frac{t\Ksf}{\Asf(\Ksf-\ell)}>1-\frac{t}{\Asf-\ell}$ when $\Ksf>\Asf$; hence $q_i^{(\Ksf)}(\Ksf t/\Asf)\geq q_i^{(\Asf)}(t)$, strictly for $i=2$.

(C\ref{cor:penalty}.II): On $[0,\Nsf/\Ksf]$ the load is $\rsf-\frac{\rsf(\rsf+1)}{2\Nsf}\Msf$, and on $[\Nsf(\Ksf-1)/\Ksf,\Nsf]$ it is $1-\Msf/\Nsf$, for every $\Ksf$.

\subsection{Proof of Corollary~\ref{cor:unknownA}}
The MAN placement does not depend on $\asf$, YMA+ attains $c_t$ with $\rsf=\min(\asf,\Nsf)$, and the abscissas $\Nsf t/\Ksf$ of the corner points of $\hat c$ do not depend on $\asf$.

\subsection{Proof of Theorem~\ref{thm:RS}}
Since the classical scheme is correct for every realization of the library, running $m$ instances in parallel we may assume that every row is a vector over $\FF_\qsf$, $\qsf=\qsf_0^m\geq\Ksf-1$. The vectors $\gv_k$ in~\eqref{eq:ex g} are the columns of the generator matrix of a $[\Ksf,\Asf]$ (extended) Reed--Solomon code, so that any $\Asf$ of them are linearly independent: for $\Asf$ vectors with $k<\Ksf$ the determinant is a Vandermonde determinant, and if $\gv_\Ksf$ is included, expanding along it leaves an $(\Asf-1)\times(\Asf-1)$ Vandermonde determinant.

\emph{Placement.} The coded row $k$ of file $n$ is $\tilde F_{n,k}:=\sum_{j\in[\Asf]}g_{k,j}F_{n,j}$, and user $k$ stores $Z_k:=\phi(\tilde F_{1,k},\ldots,\tilde F_{\Nsf,k})$, of the same size as the classical cache.

\emph{Delivery.} Given $\Ic=\{i_1,\ldots,i_\Asf\}$, the library $F'_n:=(\tilde F_{n,i_1},\ldots,\tilde F_{n,i_\Asf})=\Gm_\Ic F_n$, where $\Gm_\Ic$ acts on the rows of $F_n$, consists of independent and uniformly distributed files, since $\Gm_\Ic$ is invertible. Then $Z_{i_j}=\phi(F'_{1,j},\ldots,F'_{\Nsf,j})$ is the cache of classical user $j$ for the library $(F'_n)_n$. The server sends $(\Ic,\dv[\Ic])$ and the classical transmission for the library $(F'_n)_n$ and the classical demands $(d_{i_j})_{j\in[\Asf]}$, of load $\Rsf$.

\emph{Decoding.} User $i_j$ runs the decoder of classical user $j$, obtains $F'_{d_{i_j}}$, and computes $F_{d_{i_j}}=\Gm_\Ic^{-1}F'_{d_{i_j}}$. The map $F_n\mapsto F'_n$ acts on each file separately, which is why the demanded file can be recovered.

\subsection{Properties of RWP}

\begin{lem}[Symmetrization]
\label{lem:sym}
Every RWP scheme for $\Asf$ users, with possibly different cache functions $\phi_1,\ldots,\phi_\Asf$, can be converted into a RWP scheme with the same cache function at every user, and the same memory and load.
\end{lem}
\begin{proof}
Split every row subfile into $\Asf$ pieces, $F_{n,a}=(F^{(r)}_{n,a})_{r\in[\Asf]}$, and let every user store $\Phi(y^{(1)},\ldots,y^{(\Asf)}):=(\phi_r(y^{(r)}))_{r\in[\Asf]}$ of its own row. For $c\in[\Asf]$, run an instance of the original scheme in which user $a$ plays the role of user $c+a \pmod \Asf$, on the pieces $F^{(c+a)}_{n,a}$; within each instance the roles are distinct, so every instance is a copy of the original scheme, with $\frac1\Asf$ of the file size.
\end{proof}

\begin{lem}
\label{lem:lem-RWP}
For the RWP, \eqref{eq:lem3} holds, and
\begin{align}
H(Z_\Bc |  F_\Sc)=|\Bc|\,h_{|\Sc|},\quad\forall\Bc\subseteq[\Asf], \ \forall\Sc\subseteq[\Nsf].
\label{eq:lem1}
\end{align}
\end{lem}
\begin{proof}
Fix $k\in[\Asf]$ and $\Sc\subseteq[\Nsf]$ with $|\Sc|=s$, and let $F_{\Sc,k}:=(F_{n,k}:n\in\Sc)$. Since $Z_k$ is a function of $(F_{n,k}:n\in[\Nsf])$ and all row subfiles are independent, $(Z_k,F_{\Sc,k})$ is independent of $(F_{n,j}:n\in\Sc,\ j\neq k)$; hence
\begin{align}
h_s=H(Z_k\mid F_\Sc)=H(Z_k\mid F_{\Sc,k}).
\label{eq:rowcond}
\end{align}

\emph{\bf Upper bound in~\eqref{eq:lem3}.} Conditioning reduces entropy, so $h_s\leq H(Z_k)=h_0$. Moreover, given $F_\Sc$, $Z_k$ is a function of $(F_{n,k}:n\notin\Sc)$, which is independent of $F_\Sc$; hence $h_s\leq H(F_{n,k}:n\notin\Sc)=\frac{\Nsf-s}{\Asf}$.

\emph{Lower bound in~\eqref{eq:lem3}.} Clearly $h_s\geq0$, and by~\eqref{eq:rowcond}, $h_0=H(Z_k)\leq H(Z_k,F_{\Sc,k})=H(F_{\Sc,k})+H(Z_k\mid F_{\Sc,k})=\frac{s}{\Asf}+h_s$.

\emph{\bf Equality~\eqref{eq:lem1}.} Given $F_\Sc$, the caches are functions of distinct rows of the files not in $\Sc$, which are independent.
\end{proof}

\subsection{Proof of Theorem~\ref{thm:chain}}
Fix $L\in[\rsf]$, distinct users $u_1,\ldots,u_L$ requesting distinct files $f_1,\ldots,f_L$, and let $F_{<i}:=(F_{f_\ell})_{\ell<i}$, $Z_{<i}:=(Z_{u_\ell})_{\ell<i}$, $F:=F_{<L+1}$, $Z:=Z_{<L+1}$. Then $\Rsf\geq H(X) \geq I(X;F,Z)=H(F,Z)-H(F,Z |  X)$, with $H(F,Z)=L+Lh_L$ by~\eqref{eq:lem1}, and
$
H(F,Z |  X)=\sum_{i\in[L]}H(Z_{u_i} |  X,Z_{<i},F_{<i})
\leq\sum_{i\in[L]}H(Z_{u_i} |  Z_{<i},F_{<i})=\sum_{i\in[L]}h_{i-1},
$
where the equality uses the chain rule and $H(F_{f_i} |  X,Z_{u_i})=0$, and the last step uses~\eqref{eq:lem1}. The resulting bound holds for every $L$, and the entropies of the symmetrized scheme belong to $\Hc(\Msf)$ by~\eqref{eq:lem2} and Lemma~\ref{lem:lem-RWP}.

\subsection{Proof of Corollary~\ref{cor:closed}}
With $L=\Nsf$ and $h_\Nsf=0$, the bracket in~\eqref{eq:chain} is $\Nsf-\sum_{i=0}^{\Nsf-1}h_i$, and $h_i\leq\min(\Msf,\frac{\Nsf-i}{\Asf})$ by~\eqref{eq:lem2} and~\eqref{eq:lem3}; substituting $j=\Nsf-i$ gives~\eqref{eq:closed}. With $q:=\lfloor\Asf\Msf\rfloor$, the sum equals $\frac{q(q+1)}{2}+(\Nsf-q)\Asf\Msf$, which gives the corner points. The bound is the exact value of~\eqref{eq:chain}: $h_s=\min(\Msf,\frac{\Nsf-s}{\Asf})$ belongs to $\Hc(\Msf)$, and, writing the bracket as $\sum_{i\in[L]}(1-i\,\delta_i)$ with $\delta_i:=h_{i-1}-h_i\leq\frac1\Asf$, each term is non-negative for $i\leq\Nsf\leq\Asf$, so that the maximum over $L$ is attained at $L=\Nsf$.

\subsection{Proof of Corollary~\ref{cor:tight}}
(C\ref{cor:tight}.I): For $q=0$ and $q=1$, \eqref{eq:closed} gives the two lines. They are attained by memory sharing among $(0,\Nsf)$, the CFL point $(\frac1\Asf,\Nsf-\frac\Nsf\Asf)$, and the GV point with $g=\Nsf-1$, $(\frac{\Nsf}{\Asf(\Nsf-1)},\Nsf-\frac{\Nsf+1}{\Asf})$, which lies on the second line, with $\Asf\Msf=\frac{\Nsf}{\Nsf-1}\in(1,2]$.

(C\ref{cor:tight}.II): For $\Nsf=2$, $\frac{\Nsf}{\Asf(\Nsf-1)}=\frac2\Asf=\frac\Nsf\Asf$, so item~(C\ref{cor:tight}.I) covers the whole range.

(C\ref{cor:tight}.III): Take $L=\rsf$ in~\eqref{eq:chain} and let $\delta_i:=h_{i-1}-h_i$. Telescoping gives
$
L(1+h_L)-\sum_{i=0}^{L-1}h_i=\sum_{i\in[L]}(1-i\,\delta_i).
$
By concavity, $\delta_i$ is non-decreasing in $i$; with $h_\Nsf=0$ and $h_{\rsf-1}\leq\frac{\Nsf-\rsf+1}{\Asf}$ by~\eqref{eq:lem3}, for every $i\leq\rsf$,
$
\delta_i\leq\delta_\rsf\leq\frac{\delta_\rsf+\cdots+\delta_\Nsf}{\Nsf-\rsf+1}=\frac{h_{\rsf-1}}{\Nsf-\rsf+1}\leq\frac1\Asf.
$
Hence $\Rsf\geq\rsf-\frac1\Asf\sum_{i\in[\rsf]}i$, i.e.,~\eqref{eq:flat}, for every $h\in\Hc(\Msf)$. The YMA point with $t=1$ attains it at $\Msf=\Nsf/\Asf$.

\subsection{Proof of Corollary~\ref{cor:hotplugrow}}
By Theorem~\ref{thm:RS}, every RWP load is achievable in the hotplug model for every $\Ksf\geq\Asf$. Conversely, in a hotplug scheme obtained by embedding an RWP, the active users form a classical RWP scheme for the library $(\Gm_\Ic F_n)_n$, whose files are again independent and uniformly distributed; hence its load is at least $\Rsf^\star_{\row(\Asf,\Nsf)}(\Msf)$, for every $\Ksf\geq\Asf$.

%RWP optima N=A=3

\subsection{Proof of Theorem~\ref{thm:N=A=3}}
We use the following notation.
Files $A,B,C$ have unit size; row $k$ of file $A$ is $A_k$, of size $1/3$, and $B_1C_1$ denotes $(B_1,C_1)$. User $k$ caches $Z_k$, a function of $(A_k,B_k,C_k)$, and $h_s:=H(Z_k | s\text{ files})$, so $h_0\leq\Msf$. Consider the demand in which users $1,2,3$ request $A,B,C$, with transmission $X$, $H(X)\leq\Rsf$. By time sharing over the joint permutations of users/rows and files, we may assume that every entropy below is invariant under these permutations. The facts used are:
\begin{itemize}
\item decoding: $A$ is a function of $(X,Z_1)$, $B$ of $(X,Z_2)$, $C$ of $(X,Z_3)$;
\item RWP identities (caches depend only on their own row, rows are independent):
\begin{align}
   & H(Z_3 | A,B)=H(Z_3 | A,B_3)=h_2,\quad
\\& H(Z_2 | A)=h_1,\quad
\\& H(Z_3 | A_2)=h_0,\quad
\\& H(B_1C_1 | Z_1,A_1) = H(Z_1,A_1,B_1C_1) - H(Z_1,A_1)
\\&\qquad\qquad\qquad\ \ = 1-(\tfrac13+h_1)=\tfrac23-h_1,\quad
\\& H(Z_1,A,B,C_2)=\tfrac73+h_2;
\end{align}
%\item conditioning reduces entropy.
\end{itemize}
Only $X$, one cache, whole files and one row of at most two files appear.
Let
\begin{align}
a&:=H(X,Z_1)=H(X,Z_1,A),
\\
P&:=H(X,Z_1,A,B,C_2)\geq H(Z_1,A,B,C_2)=\tfrac73+h_2.
\end{align}
We bound $a$ in three ways, each obtained by completing $(X,Z_1,A)$ differently.

\smallskip\noindent\emph{(i) Through user 2.}
Since $H(Z_2 | X,A)\leq H(Z_2 | A)=h_1$ and $B$ is decoded from $(X,Z_2)$,
\begin{align}
a&\geq H(X,A)\geq H(X,Z_2,A,B)-h_1
\\&=H(X,Z_1,A,B)-h_1,
\end{align}
where the last step swaps users $1,2$ (and files $A,B$).
Next, $H(C_2 | X,Z_1,A,B)\leq H(C_2 | X,A,B)$ gives $H(X,Z_1,A,B)\geq P-H(X,A,B,C_2)+H(X,A,B)$, and $H(X,A,B)\geq H(X,Z_3,A,B)-h_2=3-h_2$. Therefore
\begin{align}
a\geq \tfrac{16}{3}-h_1-H(X,A,B,C_2). \tag{i}
\end{align}

\smallskip\noindent\emph{(ii) Through row 2.} $H(B,C_2 | X,Z_1,A)\leq H(B,C_2 | X,A_2)$ gives $a\geq P-H(X,B,A_2C_2)+H(X,A_2)$. Moreover, $C$ is decoded from $(X,Z_3)$, so
\begin{align}
H(X,A_2)&\geq H(X,Z_3,C,A_2)-h_0
\\&=H(X,Z_1,A,B_3)-h_0\geq H(X,A,B_3)-h_0,
\end{align}
where the equality applies the cyclic shift $1\to2\to3\to1$ of users/rows with $C\to A$, $A\to B$. Therefore
\begin{align}
a\geq \tfrac73+h_2-h_0+H(X,A,B_3)-H(X,B,A_2C_2). \tag{ii}
\end{align}

\smallskip\noindent\emph{(iii) Through row 1.} $H(B_1C_1 | X,Z_1,A)\leq H(B_1C_1 | Z_1,A_1)=\frac23-h_1$, hence, swapping users $1,2$,
\begin{align}
a\geq H(X,A,B_1C_1)-\tfrac23+h_1=H(X,B,A_2C_2)-\tfrac23+h_1. \tag{iii}
\end{align}

\smallskip\noindent\emph{Summing.} The terms $h_1$ and $H(X,B,A_2C_2)$ cancel, and
\begin{align}
3a\geq 7-h_0+h_2+H(X,A,B_3)-H(X,A,B,C_2).
\end{align}
Swapping users $2,3$ (and files $B,C$), and using that $(X,Z_3,A,B_3)$ determines $C$ and conversely $(A,C,B_3)$ determines $Z_3$,
\begin{align}
H(X,A,B,C_2)&=H(X,A,C,B_3)=H(X,Z_3,A,B_3)
\\&\leq H(X,A,B_3)+h_2 .
\end{align}
Hence $3a+h_0\geq7$. Finally $\Rsf\geq H(X)\geq H(X,Z_1)-H(Z_1)=a-h_0$, so $3\Rsf+4h_0\geq3a+h_0\geq7$, and $h_0\leq\Msf$.

%\begin{rem}
The proof is the sum, with multipliers, of $14$ Shannon inequalities (a sparse dual solution of the genie LP restricted to single-row genies); the three bounds (i)--(iii) are the three ways the LP completes $(X,Z_1,A)$ to the whole library. No pair of caches and no two-row genie is needed.
%\end{rem}

\end{document}